\documentclass[11pt]{article}
\usepackage[margin=1in]{geometry}
\usepackage{amsmath,amssymb}
\usepackage{booktabs}
\usepackage{graphicx}
\usepackage{siunitx}
\usepackage{hyperref}
\usepackage[authoryear]{natbib}
\usepackage{microtype}
\usepackage{authblk}
\usepackage{enumitem}
\usepackage{comment}
\usepackage{titlesec}
\usepackage{caption}
\usepackage{setspace}
\usepackage{longtable}
\usepackage{algorithm}
\usepackage{amsmath}
\usepackage{algpseudocode}
\usepackage{amsfonts}
\usepackage{float}

\titleformat{\section}{\large\bfseries}{\thesection}{0.75em}{}
\titleformat{\subsection}{\normalsize\bfseries}{\thesubsection}{0.75em}{}
\titleformat{\subsubsection}{\normalsize\itshape}{\thesubsubsection}{0.75em}{}

\title{Human Digital Twin: Data, Models, Applications, and Challenges}

\author[1]{Rong Pan}
\author[2]{Hongyue Sun}
\author[3]{Xiaoyu Chen}
\author[1]{Giulia Pedrielli}
\author[4]{Jiapeng Huang}
\affil[1]{School of Computing and Augmented Intelligence, Arizona State University, Tempe, Arizona, USA.}
\affil[2]{University of Georgia, Athens, Georgia, USA.}
\affil[3]{University at Buffalo, Buffalo, New York, USA.}
\affil[4]{School of Medicine, University of Louisville, Louisville, Kentucky, USA.}
\date{\today}

\begin{document}
\maketitle

\begin{abstract}
Human digital twins (HDTs) are dynamic, data-driven virtual representations of individuals, continuously updated with multimodal data to simulate, monitor, and predict health trajectories. By integrating clinical, physiological, behavioral, and environmental inputs, HDTs enable personalized diagnostics, treatment planning, and anomaly detection. This paper reviews current approaches to HDT modeling, with a focus on statistical and machine learning techniques, including recent advances in anomaly detection and failure prediction. It also discusses data integration, computational methods, and ethical, technological, and regulatory challenges in deploying HDTs for precision healthcare.
\end{abstract}


\begin{figure}[h]
  \includegraphics[width=\textwidth]{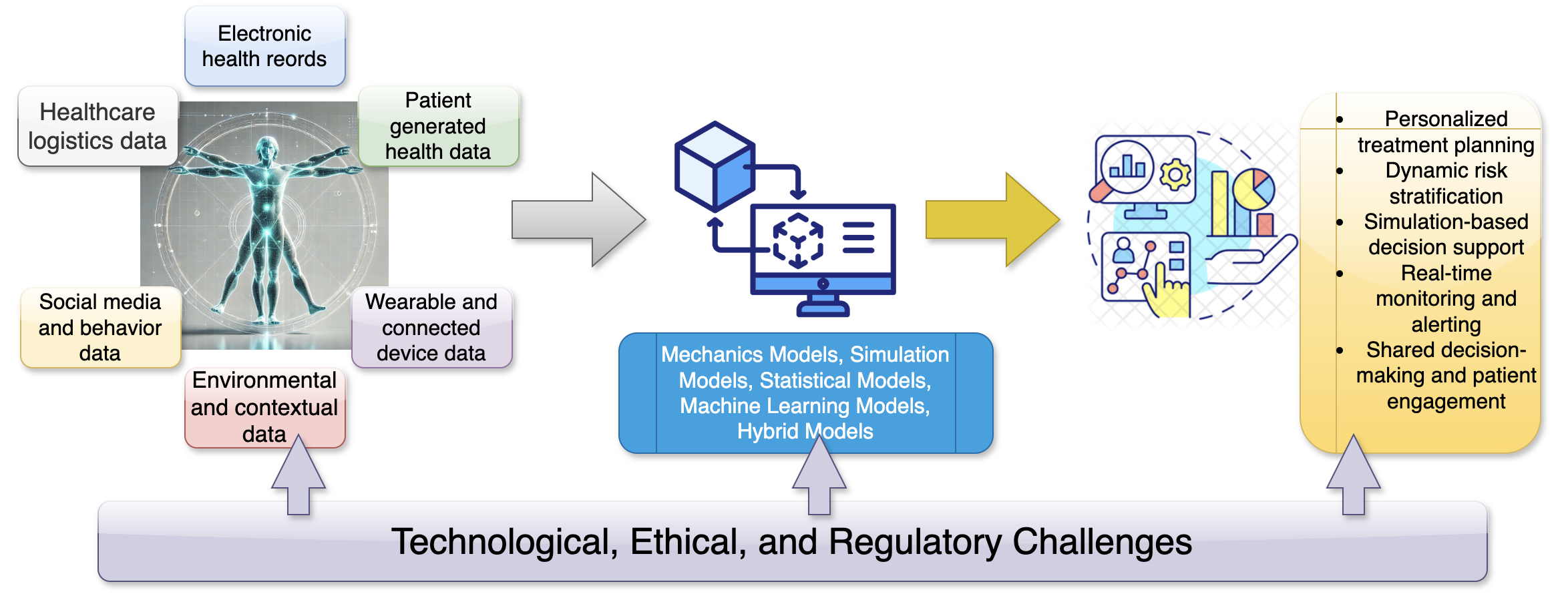}
  \caption{A Framework for Developing Human Digital Twins in Healthcare Domain.}
  \label{fig:framework}
\end{figure}

\section{Introduction}

A digital twin (DT) for human health is a dynamic computational model that integrates clinical, physiological, behavioral, environmental, and biological data to provide a personalized, evolving representation of an individual’s health throughout their ongoing care.  This DT is continuously updated with real-time or near-real-time data streams, allowing it to simulate and project the individual’s current and future health states, detect anomalies,  test treatment strategies, and guide more timely and accurate decision-making with better accommodation of individual baselines. Originally conceived in engineering as a tool to monitor and optimize complex physical systems, the DT concept is now being adapted to medicine, with the human digital twin (HDT) emerging as a transformative tool in precision health \cite{lonsdale2022perioperative}. An HDT can facilitate computer experimentation, enabling clinicians to assess, predict, and simulate a patient's responses without direct interventions on the patient’s body. It acts as a holistic, physics-guided, data-driven model that evolves alongside the patient, mirroring their unique health trajectory.

The shift toward HDTs reflects a broader movement from population-based medicine to personalized precision healthcare. Historically, clinical decisions were guided by generalized protocols derived from population averages. In contrast, an HDT makes personalized care possible by incorporating real-time multimodal data streams -- ranging widely from electronic health records (EHRs) and physiological data warehouse (PDW) to Internet of medical things (IoMT), such as wearable device outputs -- into the human health simulation and medical decision-making process. This multidimensional modeling enables clinicians to detect early signs of deterioration, predict adverse events, and tailor therapies, educates medical students about proper care methods, and engages patients in self-care strategies grounded in personalized feedback \cite{chen2023networking, vallee2024envisioning, chen2024generative, rovati2024development}. In perioperative care, for example, Lonsdale et al. \cite{lonsdale2022perioperative} describe the use of HDTs to simulate anesthetic effects, guide fluid management intraoperatively, and monitor recovery post-surgery by comparing actual versus expected recovery trajectories. Other recent surveys on this topic can be found in \cite{basu2024human, barresi2023digital, ghatti2023digital, maizi2024digital}. In this paper, we will provide an HDT overview with an emphasis on statistical and machine learning models for health prediction and anomaly detection. While there are other HDT research streams focusing on 3D models of human organs, operation room animation, and surgical robot creation, for brevity, they are not included in this workshop paper. Interested readers may refer these topics to, e.g., \cite{cochand2023systems, diniz2025digital, grob2024health, hein2024creating, herman2024human, lonsdale2023machine, martinez2019cardio, naplekov2018methods, saggiohuman}.

The current HDT developments for personalized healthcare can be characterized by four major stages:

\begin{itemize}
    \item Data consolidation and baseline model creation: Data are essential for constructing workable HDT models. The available data may come from EHRs, wearable sensors, and medical imaging, and they are used to establish a baseline model of the individual’s physiological and clinical profile.
    \item Real-time monitoring and update: With the incorporation of streaming data (e.g., continuous heart rate, activity levels), the twin becomes a dynamic mirror of the individual’s real-time health status, capable of tracking deviations from personalized baselines.
    \item Predictive simulation and intervention planning: Leveraging machine learning and hybrid physiological models, the HDT can forecast disease progression, simulate interventions, and suggest optimal treatment pathways based on individualized predictions.
    \item Clinical integration and closed-loop decision support: In its mature form, the DT functions as a clinical decision support system, offering actionable insights, triggering alerts, and guiding interventions in an adaptive, ongoing feedback loop.
\end{itemize}

To explore the current landscape and future directions of digital twins in human health, this paper will discuss their four central capabilities: modeling, monitoring, prediction, and anomaly detection. We will review the data sources required to build and maintain an HDT, the modeling and simulation methods that enable its functions and real-world applications, and the technical, ethical, and regulatory challenges that must be addressed to translate digital twin technology into clinical practice.

\section{Data for Building HDTs}

The foundation of an HDT lies in its capacity to integrate diverse and multimodal data streams into a unified, continuously evolving representation of an individual’s health. This integration spans structured clinical data, physiological signals, molecular profiles, environmental exposures, and patient-reported experiences \cite{lonsdale2022perioperative}. For example, EHRs can provide an individual's clinical history for a digital twin, including demographics, diagnoses, medication history, laboratory results, procedures, and clinical notes. These data enable the twin to define baseline health conditions, identify comorbidities, and assess historical treatment outcomes. These individual health records may also include genomic and molecular data, and medical imaging, including MRI, CT, and ultrasound, which can contribute to the anatomical fidelity of HDT. These data modalities provide additional layers of personalization by enabling the digital twin to incorporate information about inherited traits and gene expression profiles, or patient-specific geometries for constructing biophysical simulations of organs and tissues. In cardiovascular applications, for example, imaging-derived vessel morphology can be integrated into fluid dynamics models to simulate blood flow under different physiological or pathological conditions \cite{corral2020digital}. This combination of imaging and computational modeling enhances the ability of an HDT to support surgical planning, device selection, or hemodynamic risk assessment.

The medical data discussed above are extremely important for building HDTs, but they are typically episodic, documenting health states during clinical encounters rather than continuously. This limitation highlights the need for integrating additional sources that reflect real-time and longitudinal health behaviors \cite{bruynseels2018digital}. To complement clinical data, patient-generated health data (PGHD) can enrich HDT by incorporating subjective and behavioral insights into the model. These data may include self-reported pain scores, mood, dietary habits, or symptom diaries recorded via mobile health applications \cite{jim2020innovations}. When combined with physiological monitoring data from wearables and connected devices, PGHD enables the twin to understand how lifestyle, behavior, and symptom variability interact over time. Note that these sensor data provide high-frequency, real-world signals such as heart rate, oxygen saturation, activity levels, and sleep quality. These data streams enable the twin to mirror the person’s moment-to-moment physiological state and learn individual baselines over time. For example, changes in heart rate variability detected by a wearable device may signal autonomic stress or inflammation before overt symptoms arise \cite{jafleh2024role}. This capability is particularly relevant in perioperative care, where digital twins can incorporate intraoperative and postoperative vital sign trajectories to guide personalized recovery pathways.

Beyond individual physiology, environmental and contextual data, collectively termed the exposome, also play an important role in determining health outcomes. Variables such as air quality, temperature, humidity, noise exposure, and geographic location influence the onset and progression of diseases such as asthma, cardiovascular diseases, and even mental health disorders. Integrating environmental data allows the HDT to contextualize physiological or behavioral deviations, reducing false alarms and improving predictive accuracy \cite{juarez2020use}. 

Finally, there is growing interest in digital behavioral data such as social media activity, smartphone usage, and online interaction patterns, which may reflect early signs of mental health changes or cognitive decline. While still experimental, studies have shown that passive digital traces can be correlated with well-being indicators such as mood, social connectivity, and stress \cite{settanni2015sharing}. With appropriate ethical safeguards, these data streams could expand the HDT awareness of psychosocial factors and support more holistic health monitoring.

To merge these diverse data sources into a functioning HDT, robust data fusion and management techniques are required. This includes aligning data temporally, standardizing units and semantics, and managing missing or inconsistent inputs. While the technical infrastructure for real-time integration is still evolving, cloud-based systems, standardized health data APIs, and privacy-preserving pipelines are helping to bridge gaps between traditionally siloed datasets \cite{lonsdale2022perioperative, jafleh2024role, juarez2020use}. By combining biological, physiological, behavioral, and environmental data into a coherent and dynamic model, HDTs are uniquely positioned to enable precise, personalized, and predictive healthcare.

\vspace{-6pt}
\section{Computational Models of HDTs}
The core of an HDT is its computational model, an executable mathematical object that transforms diverse data inputs into predictions of human health states. In general, HDTs employ three principal modeling strategies: mechanistic (physics-based), data-driven (statistical and machine learning-based), and increasingly, hybrid approaches that combine both.

Mechanistic or physics-based models use mathematical representations of physiological systems derived from established biological and physical principles \cite{coveney2005modelling}. These models, often structured as systems of differential equations, describe core functions such as cardiac hemodynamics \cite{koivumaki2022computational}, metabolic regulation \cite{espinel2024hybrid}, or respiratory mechanics \cite{warnaar2023computational}. For example, cardiovascular twins frequently simulate pressure and flow using lumped-parameter models of the circulatory system or computational fluid dynamics (CFD) based on patient-specific vessel geometry from imaging data \cite{chakshu2021towards}. These models are prized for their physiological fidelity and interpretability, especially in domains like cardiology and surgery where cause-and-effect understanding is critical. However, mechanistic models often require high-fidelity human organ and biological system geometries, and their computational cost increases sharply with complexity and scale. 

Discrete event simulation (DES) models are another set of hospital logistical (e.g., surgical operations) process-based models. When integrated into a digital twin of the operating room (OR), DES provides a flexible and interpretable way to simulate, evaluate, and optimize surgical workflows and perioperative operations. Its ability to model resource constraints explicitly, including limited surgical teams, equipment availability, or room turnover capacity, makes it ideal for scenario testing and "what-if" analyses, which is particularly valuable for hospital administrators.

Another type of HDT model is data-driven models, including both statistical models and machine learning (ML) models. Statistical models offer a mathematically grounded and often computationally efficient framework for analyzing uncertainty, estimating parameters, and generating probabilistic predictions from clinical and physiological data. In addition, statistical approaches explicitly characterize data distributions and covariate relationships, which is essential in high-stakes applications like healthcare. Recently, as highlighted in Section 3.2, there has been growing interest in non-Euclidean statistical models, such as the distribution-in-distribution-out (DIDO) regression, which works on data represented as probability distributions rather than point values \cite{chen2024distribution}. This approach is particularly suited for health data that is naturally stochastic and continuous, such as vital sign readings during surgery.

Also in recent years, ML-based models have gained momentum as a way to learn patterns directly from large-scale datasets. These models can forecast patient trajectories (e.g., disease onset or deterioration) and personalize care by learning from population-level data. In a similar context, these ML-models have been used to model, predict and assess the human fatigue and risks during operational tasks \cite{hajifar2021forecasting,lamooki2022data,kheiri2023functional,kheiri2023human} and organ transplantation \cite{hajifar2022online,hajifar2024irregular}. In the digital twin context, ML enables adaptive personalization, where the model evolves with each new data point. These data-driven models also underpin many of the anomaly detection and failure prediction methodologies, as will be discussed in Section 4. Methods such as Gaussian Process regression, change-point detection, and survival analysis (e.g., Cox proportional hazards models) allow HDTs to detect regime shifts, model degradation trajectories, and estimate the remaining useful life (RUL) of a physiological system.

Integrating mechanistic and data-driven components, the hybrid modeling approach offers a promising pathway to leverage the strengths of both. In this approach, physiological knowledge constrains or informs the learning process of ML models. A prime example is physics-informed neural networks (PINNs), where differential equations representing known physiology are incorporated into the training loss of the neural network, ensuring outputs remain physiologically plausible \cite{raissi2019physics}. Thus, ML models may serve as surrogates for computationally intensive simulations, accelerating prediction without sacrificing accuracy \cite{sharma2023physics}. Such hybrid strategies help balance accuracy, efficiency, and interoperability requirements for clinical trust and adoption.

\vspace{-6pt}
\subsection{DIDO Regression: A Multiple Regression Model Operating with Distribution-valued Samples}
Multiple linear regression models have been the gold standard to facilitate knowledge discovery in biomedical and healthcare domains due to their simplicity for interpretation and inference. 
Such regression models assume that both predictors and responses are scalar-valued and lie in an Euclidean space with well-defined arithmetic operations. However, many physiological signals (such as blood pressure, heart rate, or glucose levels) are measured with non-negligible level of uncertainty due to measurement errors and underlying auto-regulation (e.g., Baroreceptor Reflex) and are best represented as probability distributions, which are elements of the Wasserstein space. Equipped with a 2-Wasserstein metric, the $L^2$-Wasserstein space was shown to be a non-flat pseudo Riemannian manifold in Otto's seminal work \cite{Otto31012001}. The technical challenge of defining linear additive structures between distributions lies in the ill-defined arithmetic operations. The DIDO regression defined the addition and scalar multiplication in $L^2$-Wasserstein space between logarithm maps based on the Levi-Civita connection between tangent spaces. Such arithmetic operations enabled the construction of linear additive structures, making it possible to define and estimate a multiple linear regression model that is both interpretable and statistically sound.

The key to the usefulness of DIDO regression in healthcare is its ability to accommodate uncertainty in both inputs and outputs, reflecting the real-world noise and biological variability inherent in physiological systems. The model yields a Fréchet least squares estimator that generalizes the classic Gauss-Markov theorem to $L^2$-Wasserstein space, ensuring that the estimator remains the best linear unbiased estimator (BLUE) under relaxed \textit{i.i.d.} assumptions. Furthermore, in the special case where all input and output distributions are univariate Gaussian, the model admits closed-form solutions for regression coefficients, predicted distributions, and explained variance, greatly improving computational efficiency. In \cite{chen2024distribution}, the DIDO regression has been used to predict intraoperative cardiac output distributions from a set of distribution-valued vital signs (e.g., distributions of blood pressure and heart rate). This approach outperforms traditional linear regression models in high-noise, small-sample settings, which makes it a promising tool for personalized precision healthcare.

\section{Monitoring, Prediction, and Change Detection}

One of the most transformative capabilities of an HDT is its ability to serve as a continuous health sentinel, reflecting an individual’s physiological state in real-time, anticipating changes, and identifying anomalies before they become critical. Through seamless integration of data, models, and feedback mechanisms, the HDT offers a proactive approach to care: shifting the clinical paradigm from episodic intervention to ongoing surveillance, prediction, and early response. However, to realize these applications, it is crucial to maintain synchronization between physical and digital twins by regularly assimilating physiological and behavioral data from sensors, wearables, and clinical systems. An HDT also develops a personalized baseline derived from an individual’s unique patterns of physiological activity over time. By modeling these personalized patterns, the HDT establishes what is “normal” for a specific person, not what is typical for the population. This context-aware monitoring allows the twin to detect subtle deviations that would be missed by standard alarms, especially in patients with atypical presentations or chronic conditions \cite{rivera2019towards, volkov2021digital, kumari2025sensors}. In this section, we highlight some recent developments in ML-based process change detection.

\vspace{-6pt}
\subsection{Change Detection: A Machine Learning Approach}

Recent advances in change point detection (CPD) and predictive modeling offer powerful tools to enhance the anomaly detection capabilities of HDTs. These methods allow HDTs not only to identify unusual deviations in physiological data but also to anticipate potential health crises before they occur.

Derivative-aware change detection (DACD)
is a novel anomaly detection method that leverages Gaussian processes (GPs) and their derivative processes to identify abrupt shifts in temporal data \cite{zhao2023active}. The key insight is that changes in the underlying health state manifest as peaks or irregularities in the first-order derivative of the GP model. By modeling both the function and its derivatives, DACD can detect sharp transitions in time series data, such as a sudden drop in oxygen saturation or an irregular heart rate spike, without making strong parametric assumptions about the data's distribution. For more complex or nonstationary data, such as long-term rehabilitation metrics or chronic disease patterns, deep Gaussian processes (DGPs) offer a flexible and powerful modeling approach \cite{zhao2025dgp}. In this method, the DGP can be trained on the patient’s time series data and transformed into the frequency domain via a sliding-window Fourier transform. This transformation allows the detection of subtle shifts in the frequency spectrum.

Beyond change detection, digital twins must also predict potential failures to enable preventive care. Based on DACD, the Gaussian derivative change point detection (GDCPD) method offers a framework for doing so by detecting multivariate change points across multiple signal channels and integrating them with a remaining useful life (RUL) prediction model. In this setup, GP derivatives are used to track subtle degradations in high-dimensional data (e.g., multichannel wearable sensor outputs), and weighted Mahalanobis distance (WMD) thresholds are established to trigger early warnings. More details of this methodology can be found in \cite{zhao2025gaussian}.

\vspace{-6pt}
\subsection{Continual Learning in the Health Context}
Continual learning (CL) develops methods capable of acquiring new knowledge and adapting to evolving data distributions over time. Such systems learn incrementally and sequentially, allowing the creation and improvement in ``expertise'' while building upon previously acquired knowledge. This is particularly crucial in real-world applications where data is dynamic, and tasks may change or expand, such as in the human system and the plethora of sensors available. CL approaches have been explored in the context of wearable computing such as in human activity recognition (HAR)~\cite{sah2022continual, lapnethar}. For example, the LAPNet-HAR framework leverages prototypical networks and a replay buffer to support adaptation~\cite{lapnethar}. These approaches have been evaluated with datasets that are specific to certain scenarios such as PAMAP2~\cite{reiss2012creating}, DSADS~\cite{altun2010comparative}, WISDM~\cite{WISDM} and SKODA~\cite{SKODA} (e.g., quality checks performed by assembly-line workers in a car maintenance scenario or data collected in laboratory settings) and do not represent free-living conditions. 

The problem of change detection is core to the literature of continual learning. Specifically, changes in the data stream are formalized as (1) domain shifts, (2) change in domain, and (3) noise in the data label. In the following we highlight the principles and key challenges in these areas.

\noindent\underline{Quantifying domain shift.} 
The challenge of data diversity in continual scenarios can be formulated as a \textit{domain shift} problem with the objective of assessing model generalizability on new data. An approach frequently employed to address domain shift is transfer learning or domain adaptation~\cite{gjoreski2019cross, 10.1145/2971763.2971764, wang2018deep, zhao2020local}. Researchers have defined changes in sensor-based data as a general distribution shift that affects data and, as a result, impacts the model performance. However, identifying the cause of the shift and understanding its influence on the data can provide a solid understanding of the nature of the distribution shift and will guide the selection of the optimal method to address the problem. 

\noindent\underline{Domain generalization} targets the problem of building models that generalize well to unseen data without access to the new training data. Still a comprehensive examination of the issue of domain generalization in relation to the various distribution shifts that emerge in pervasive CL is missing. In machine learning community, a range of domain generalization algorithms are proposed for computer vision and natural language processing tasks~\cite{wang2022generalizing}.

\noindent\underline{Handling noise in data.}
To realize continual learning in pervasive systems, an initial robust model needs to be trained using labeled data collected in uncontrolled settings. The common approach for gathering such labels is using EMA (ecological momentary assessment) methods which involve repeated sampling of the user's current behaviors and experiences (e.g., activity, stress, diet) in real-time. However, the labels expressed by end-users exhibit significant amounts of inter-user and intra-user variations leading to poor performance of the model trained on such labeled data in continual learning settings.

\section{Technological, Ethical, Regulatory Challenges}

Despite their promise to revolutionize healthcare, HDTs raise a host of challenges that span technical implementation, ethical use, and regulatory oversight. These concerns are central to the safe, scalable, and equitable deployment of HDTs in real-world medical environments. This section outlines some obstacles that must be addressed to fully realize the potential of digital twin technology in human health. Note that it is certainly not a comprehensive list.

\begin{itemize}
    \item Technological challenges: The creation and deployment of HDTs involve numerous computational and data-related hurdles. First, integration of heterogeneous, multimodal data, including time-series sensor signals, EHRs, imaging data, genomic profiles, patient-reported outcomes, etc., is a tremendously difficult task. Second, while modeling a single patient in a controlled research setting is feasible, deploying HDTs at scale demands automated pipelines for model instantiation, calibration, and updating across diverse populations. This involves not only computational efficiency but also robust methods for real-time model adaptation, such as online learning or streaming Bayesian inference. Additionally, uncertainty quantification remains an unsolved challenge. In high-stakes environments like intensive care or oncology, clinicians must know not only what the HDT predicts but how confident it is in those predictions. Models must, therefore, include calibrated probabilistic outputs and explainable uncertainty bounds, especially when using machine learning or hybrid architectures. Failure to do so risks overconfidence in untrustworthy predictions.
    \item Ethical challenges: The development and use of HDTs raise several profound ethical concerns, beginning with patient autonomy and consent and continuing with ingesting, processing, and simulating patient data. This introduces ambiguity around informed consent: when does a patient consent to updates, simulations, or interventions made by their twin? Moreover, how are patients involved in decisions made by or with the twin, and how can they contest or override its recommendations? Bias and fairness are also critical concerns. HDT models trained on biased datasets may perpetuate or even exacerbate disparities in healthcare outcomes. Last but not least, data privacy and security constitute another major ethical challenge. HDTs operate on sensitive health data, some of which may include behavioral, genomic, or social context information. This data, if compromised, could be used to infer personal details or discriminate in employment, insurance, or social settings. 
    \item Regulatory and legal challenges: The evolving landscape of HDTs has outpaced existing regulatory frameworks, which were not designed to govern continuously learning, personalized, simulation-based systems. Regulatory bodies such as the U.S. Food and Drug Administration (FDA), the European Medicines Agency (EMA), and others are grappling with how to assess and approve HDTs. While digital health tools like clinical decision support software (CDSS) and mobile health apps are subject to existing oversight, HDTs often straddle multiple categories: part medical device, part data system, part simulation engine. This makes it unclear which regulatory pathways apply. Moreover, data provenance, auditability, and traceability are central to regulatory trust. Regulators and clinical users must be able to trace the data lineage, modeling assumptions, and version histories that led to a particular output or prediction. This is essential for both transparency and post-hoc error analysis. The increasing use of machine learning models complicates this further, particularly when “black-box” models are involved.
\end{itemize}

\section{Conclusion}

HDTs represent a transformative paradigm for precision healthcare, offering the ability to model, monitor, predict, and detect anomalies in an individual’s health state. By integrating multimodal data and leveraging advanced modeling techniques, HDTs can enable personalized, real-time decision support and proactive care. However, significant technological, ethical, and regulatory challenges remain. These include data integration, model transparency, bias mitigation, and the need for adaptive, lifecycle-aware regulatory frameworks. Addressing these challenges will be essential for HDTs to achieve widespread clinical adoption.


\bibliographystyle{ACM-Reference-Format}
\bibliography{ref}

\end{document}